**REVIEW** Open Access

# Recent advances in methodology for clinical trials in small populations: the InSPiRe project

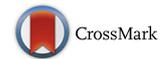

Tim Friede[1], Martin Posch[2], Sarah Zohar[3], Corinne Alberti[4], Norbert Benda[5], Emmanuelle Comets[6,7], Simon Day[8], Alex Dmitrienko[9], Alexandra Graf[2], Burak Kürsad Günhan[1], Siew Wan Hee[10], Frederike Lentz[5], Jason Madan[10], Frank Miller[11], Thomas Ondra[2], Michael Pearce[12], Christian Röver[1], Artemis Toumazi[4], Steffen Unkel[1], Moreno Ursino[3], Gernot Wassmer[2] and Nigel Stallard[10*]

## Abstract

Where there are a limited number of patients, such as in a rare disease, clinical trials in these small populations present several challenges, including statistical issues. This led to an EU FP7 call for proposals in 2013. One of the three projects funded was the Innovative Methodology for Small Populations Research (InSPiRe) project. This paper summarizes the main results of the project, which was completed in 2017.

The InSPiRe project has led to development of novel statistical methodology for clinical trials in small populations in four areas. We have explored new decision-making methods for small population clinical trials using a Bayesian decision-theoretic framework to compare costs with potential benefits, developed approaches for targeted treatment trials, enabling simultaneous identification of subgroups and confirmation of treatment effect for these patients, worked on early phase clinical trial design and on extrapolation from adult to pediatric studies, developing methods to enable use of pharmacokinetics and pharmacodynamics data, and also developed improved robust meta-analysis methods for a small number of trials to support the planning, analysis and interpretation of a trial as well as enabling extrapolation between patient groups. In addition to scientific publications, we have contributed to regulatory guidance and produced free software in order to facilitate implementation of the novel methods.

## Background

A disease is defined as rare by the European Union if the prevalence is no more than 5 per 10,000 [1], and by the United States if it affects fewer than 200,000 people in the US [2], equivalent to 62 per 100,000 in 2015 [3]. European regulatory guidance [1] states that "patients with [rare] conditions deserve the same quality, safety and efficacy in medicinal products as other patients; orphan medicinal products should therefore be submitted to the normal evaluation process". This is in agreement with United States guidance [4] that "The Orphan Drug Act [...] does not create a statutory standard for the approval of orphan drugs that is different from the standard for drugs for common conditions. Approval of all drugs – for both rare and common conditions – must be based on demonstration of substantial evidence of effectiveness in treating or preventing the condition and evidence of safety for that use". Rigorous clinical trial evaluation of treatments is thus as necessary in rare diseases as in more common ones. The European Medicines Agency acknowledges that this represents a challenge, however, indicating that "it may be that in conditions with small and very small populations, less conventional and/or less commonly seen methodological approaches may be acceptable if they help to improve the interpretability of the study results". This suggests that there is a need for development of novel methodology for the design and conduct of clinical trials and analysis of the trial outcomes in research in small patient populations. It was this need that led to a call for proposals under the European Union's Seventh Framework Programme for Research, Technological development and Demonstration (EU FP7) in 2013 for projects that would "develop new or improved statistical design methodologies for clinical trials aiming at the efficient assessment of [...] a treatment for small population groups in particular for rare diseases or personalized [...] medicine" [5]. Three projects were funded under this call; the Innovative Methodology for Small Populations Research (InSPiRe) project (www.warwick.ac.uk/inspire), the Integrated Design and Analysis of Small Populations Group Trials (IDeAl) project (www.ideal.rwth-aachen.de) and the Advances in Small Trials Design for Regulatory Innovation and Excellence (Asterix) project (www.asterix-fp7.eu) [6].

*Correspondence: n.stallard@warwick.ac.uk
[10]Warwick Medical School, University of Warwick, Coventry, UK
Full list of author information is available at the end of the article

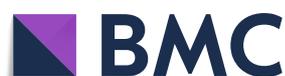

The aim of this paper is to summarize the main results of the InSPiRe project. This project, completed in 2017, brought together a team of experts from eight institutions, including academia, industry and regulatory authorities, in five European countries, with additional guidance from an Advisory Board including methodological and clinical experts and patient representatives.

Faced with the challenge of the design, conduct and analysis of clinical trials in small population groups, we have focused on a number of areas where we felt that methodological development was both needed and feasible. In particular, we have developed novel statistical methodology in the two broad areas of *efficient study design* and *improved analysis and evidence synthesis*. Efficient study design is particularly important for clinical trials in small populations as it enables the maximum information to be obtained from the sometimes necessarily limited small sample size, whilst improved analysis and evidence synthesis ensures that as much relevant information as possible is obtained and used in the analysis and interpretation of the results. This can include use of information on endpoints other than the primary endpoint in the trial as well as information from sources external to the trial, including data from other trials, observational studies and disease registries [7]. The latter can include extrapolation methods, for example, between studies in adults and children. This is an area that can be controversial, but is one where we believe further methodological and applied work is clearly justified.

In the InSPiRe project we have developed new methods in four specific areas (see Table 1), two relating to efficient design and two to improved analysis and evidence synthesis; the determination of optimal designs for confirmatory studies using decision-theoretic and value-of-information (VOI) approaches, the design of confirmatory studies with stratified populations for personalized medicines, the incorporation of pharmacokinetics (PK) and pharmacodynamics (PD) data in early-phase dose-finding studies, and meta-analysis methods for small trials or small numbers of trials. The work in these four areas is described below.

**Table 1** Main project topics and outputs

| Efficient study design | |
|---|---|
| Optimal designs for confirmatory studies using decision-theoretic and value-of-information (VOI) approaches | Design of confirmatory studies with stratified populations for personalized medicines |
| Key publications: [12–15] | Key publications: [17, 20–23] |
| **Improved analysis and evidence synthesis** | |
| Incorporation of pharmacokinetics (PK) and pharmacodynamics (PD) data in early-phase dose-finding studies | Meta-analysis methods for small trials or small numbers of trials |
| Key publications: [25, 27–30] | Key publications: [33, 34, 36–38, 40, 41] |
| Open-source R software: `dfpk` [26], `dfped` [29] | Open-source R software: `bayesmeta` [35], `nmaINLA` [42] |

## Decision-theoretic and value-of-information designs for clinical trials in small populations

Most methodology for clinical trial design makes no reference to the size of the population in which the research is conducted. Whilst this may be reasonable in a large population, in rare diseases or other small populations it could lead to designs that are inappropriate.

In order to establish the context for future research work, we completed an analysis of trials in rare diseases recorded in the ClinicalTrials.gov database as well as exploring novel methods. This showed that the sample sizes in phase 2 trials in rare diseases were similar for different prevalence but that phase 3 trials in rare diseases with lowest prevalence were statistically significant lower than those in less rare diseases and were more similar to those in phase 2 as shown in Fig. 1 [8].

We have considered determination of appropriate decision-making methods for small population clinical trials. In particular we have explored the use of a Bayesian decision-theoretic framework [9] to compare the costs of clinical trial evaluation with the potential benefits to current and future patients, assessing how the cost-benefit balance differs between large and small patient populations when in the latter patients recruited to a clinical trial could be a substantially proportion of the population. As recruitment to one clinical trial may also affect the number of patients that can be recruited to other trials when the population under investigation is small [10], we have also considered the design of a series of trials in a small population group.

We completed a systematic literature review on the use of decision-theoretic approaches in clinical trial designs, with a view to providing an overview of the current trends.

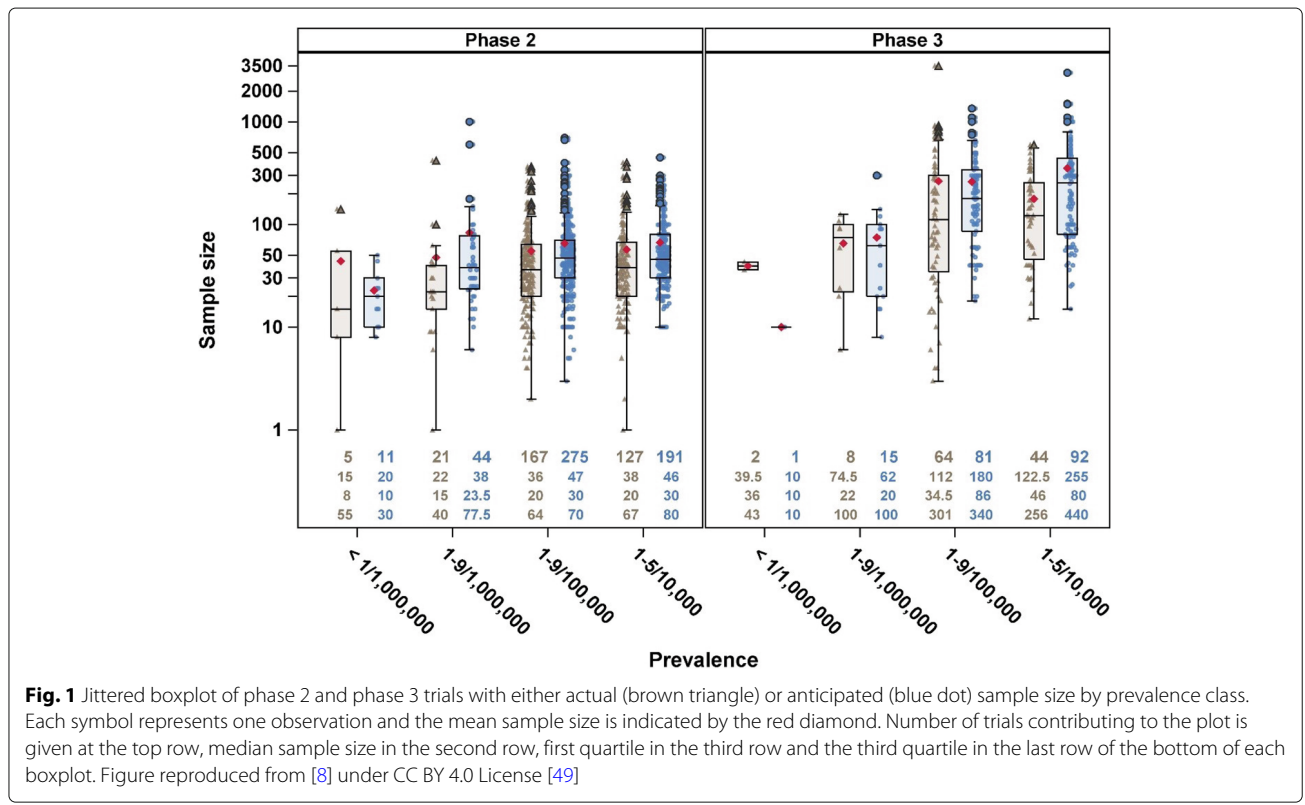

**Fig. 1** Jittered boxplot of phase 2 and phase 3 trials with either actual (brown triangle) or anticipated (blue dot) sample size by prevalence class. Each symbol represents one observation and the mean sample size is indicated by the red diamond. Number of trials contributing to the plot is given at the top row, median sample size in the second row, first quartile in the third row and the third quartile in the last row of the bottom of each boxplot. Figure reproduced from [8] under CC BY 4.0 License [49]

This systematic review identified 67 articles proposing decision-theoretic design methods relevant to small clinical trials. The review discusses these in detail, classifying them according to the type of study design and gain function proposed [11].

Building on this existing work, we have developed methodology for the use of a VOI method for a confirmatory phase III trial, particularly in the small population setting [12]. This has two important consequences in terms of optimal design; it challenges the usual method of sample size determination based on frequentist error rates and shows that in a small population setting a smaller trial than usual may be optimal.

In detail, we determined the optimal sample size and significance level for a frequentist hypothesis test at the end of a trial, and investigated how these change with the population size. We showed how decision-theoretic VOI analysis suggests a more flexible approach with both type I error rate and power (or equivalently trial sample size) depending on the size of the future population for whom the treatment under investigation is intended. Taking a more general viewpoint, we have shown that for a wide range of distributions, including those for continuous, binary or count responses, and gain function forms, the optimal trial sample size is proportional to the square root of the population size, with the constant of proportionality depending on the gain function form and prior distribution of the parameters of the distribution of the data [13].

We have compared this method with alternative sample size approaches in three case studies; Lyell's disease, adult-onset Still's disease and cystic fibrosis [14]. In each case we outline in detail the reasonable choice of parameters for the different approaches and calculate sample sizes accordingly. This work illustrates the influence of the input parameters in the different approaches and we recommend investigating different sample size approaches before deciding finally on the sample size.

We have also developed decision-theoretic methods for the simultaneous design of a series of trials in a small fixed population. Use of the methodology has been illustrated through retrospective application in an example in small orthopaedic surgery trials [15].

Further work to extend the models developed is ongoing. In particular, we are exploring the optimal design of multistage trials, settings in which the disease prevalence is considered unknown, with information obtained from the rate of recruitment to the trial itself, and designs that are optimal for more different stakeholders such as regulatory authorities and industrial sponsors.

## Research in confirmatory trials for small populations and personalized medicines

The development of targeted therapies that act on certain molecular mechanisms of diseases requires specific trial design and analytical methods. Their objective is the prediction of patients' outcomes based on genetic features or other biomarkers, to identify and confirm subgroups of patients for which the therapy's benefit risk balance is positive.

We performed a literature search to summarize the currently available methodology for the identification and confirmation of targeted subgroups in clinical trials [16]. In total 86 scientific articles proposing relevant methods were identified that were classified as confirmatory, exploratory or applicable in a confirmatory as well as exploratory settings. The review identified a wide range of trial designs, including fixed sample, group sequential, and several types of adaptive designs.

In our work we have considered designs where subgroups are defined based on a continuous biomarker and several thresholds are considered to define the subgroup.

We derived confirmatory testing procedures that control false positive rates if several thresholds are under consideration [17] and show that the type I error rate of earlier proposed testing procedures based on group sequential rejection boundaries may be inflated if the biomarker has a prognostic effect (e.g., if it is correlated with the prognosis of patients in the absence of a treatment effect). Consequently, we propose improved hypotheses testing approaches based on regression models and combination tests that robustly control the familywise error rate. We also investigated adaptive enrichment designs. In these two-stage designs, in the first stage patients are recruited from the full population. Following an interim analysis, based on the interim data, the design of the second stage may be modified. For example, recruitment may be limited to patients in a subgroup of biomarker positive patients and/or the sample sizes in the subgroups may be adapted [18].

We provided a comprehensive description of the statistical methodologies for confirmative adaptive designs with multiple objectives and their application in adaptive two-stage enrichment designs [19, 20]. For the special case of adaptive designs with a survival endpoint, hypothesis tests were developed that allow for early rejection of the null hypothesis at an interim analysis. This work generalizes earlier adaptive procedures that control the familywise type I error rate in the strong sense but have limitations in that they either cannot use information from surrogate endpoints for adaptive decision making or do not allow early rejections at an interim analysis.

To guide the design of clinical trials for the development of targeted therapies, working together with the IDeAl project, we developed a decision-theoretic framework to optimize single stage and adaptive two-stage designs [21–23]. To address the incentives of different stakeholders, we proposed utility functions representing the benefit of a particular clinical trial from a sponsor's and society's perspective. Here we assume that the utility of the sponsor is the net present value of a trial, while for society it is the expected health benefit adjusted for the trial cost. In the planning phase, expected utilities for different trial designs and different utility functions are computed based on Bayesian prior distributions for the effect sizes in the subgroup and the full population. Then optimal trial designs are identified that maximize these expected utilities by optimizing the sample size, the multiple testing procedure and the type of the design. The considered types of trials include classical designs, where no biomarker information is used and only the full population is tested, enrichment designs, where only biomarker positive patients are included, stratified designs, where patients from the full population are included and the treatment effect is tested in the subgroup and the full population, and partial enrichment designs, where the prevalence of the subgroup in the trial is a design parameter that can be chosen to maximize the expected utility.

We found that the optimal trial designs depend on the prevalence of the subgroup, the strength of the prior evidence that the treatment effect varies across subgroups, and on the cost of biomarker development and determination. Furthermore, we observe that optimal designs for the sponsor and the societal view differ. Trials optimized under the sponsor view tend to have smaller sample sizes and are conducted in the full population even in settings where there is substantial prior evidence that the treatment is effective in the subpopulation only. This is due to the fact that the variability of treatment effect estimates means a treatment might appear effective in a subpopulation (and bring a gain for the sponsor) even if it is not effective and has no benefit for patients.

We also extended the work to consider adaptive two-stage enrichment designs. We showed that adaptive enrichment designs can lead to a higher expected utility than single stage designs, especially in settings where there is high uncertainty if the treatment is effective only in a subgroup. Figure 2 illustrates the results of optimizing interim adaptation rules to maximize the expected utilities by extensive simulations and a dynamic programming algorithm. As for single stage designs, we observe differences in the optimized designs if trials are optimized under the sponsor or the societal perspective. An important advantage of adaptive designs compared to single stage designs is their increased robustness with regard to a misspecification of the planning assumptions.

## Extrapolation and use of available information in early-phase studies

Early phase dose-finding studies aim to obtain reliable information on an appropriate dose for use in further clinical trials. The designs used have generally relied primarily on observed toxicity data [24]. We have proposed novel

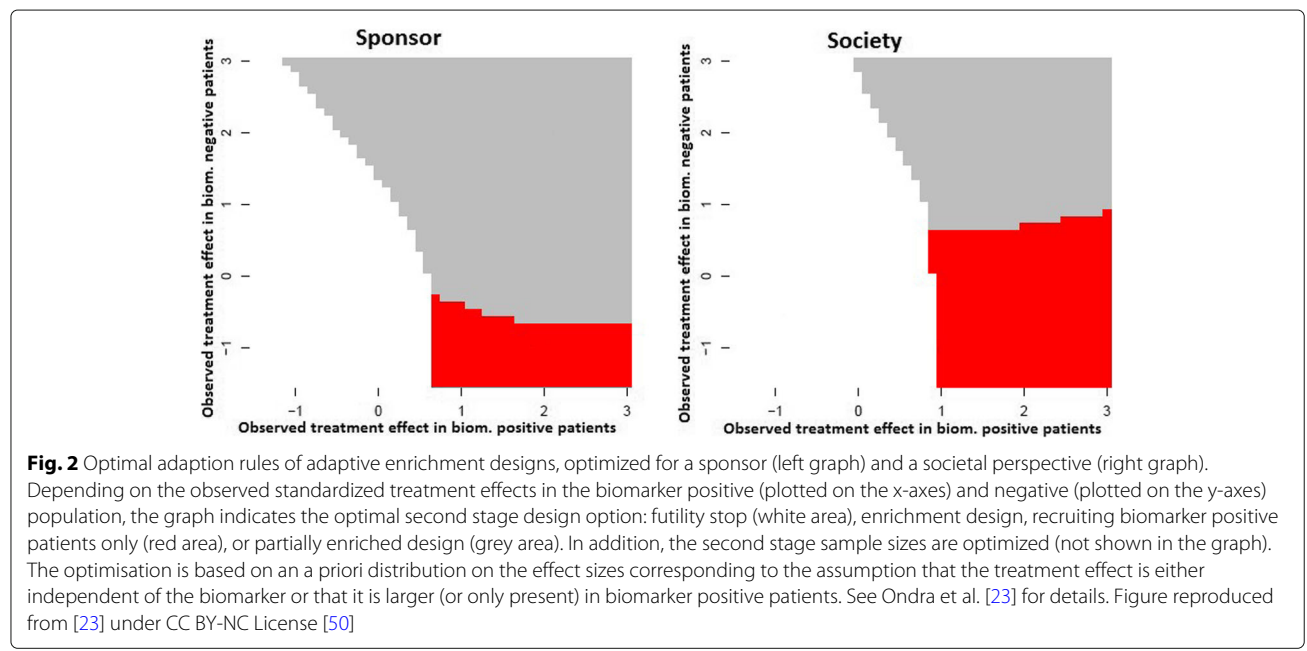

**Fig. 2** Optimal adaption rules of adaptive enrichment designs, optimized for a sponsor (left graph) and a societal perspective (right graph). Depending on the observed standardized treatment effects in the biomarker positive (plotted on the x-axes) and negative (plotted on the y-axes) population, the graph indicates the optimal second stage design option: futility stop (white area), enrichment design, recruiting biomarker positive patients only (red area), or partially enriched design (grey area). In addition, the second stage sample sizes are optimized (not shown in the graph). The optimisation is based on an a priori distribution on the effect sizes corresponding to the assumption that the treatment effect is either independent of the biomarker or that it is larger (or only present) in biomarker positive patients. See Ondra et al. [23] for details. Figure reproduced from [23] under CC BY-NC License [50]

methods for (i) incorporating of the PK/PD information in the dose-allocation process, (ii) planning and conducting clinical trial for reducing neonatal seizures for which no other method was available, (iii) proposing extrapolation methods for bridging studies from adults to children and (iv) incorporating subjective information, such as physicians' elicitation weighted by their degree of expertise, into the study design.

We proposed and compared methods to incorporate PK measures in the dose allocation process during phase I clinical trials. PK observations were incorporated in a number of different ways; as a covariate, as a dependent variable or in a hierarchical modelling approach. We conducted a large simulation study which showed that adding PK measurements as a covariate alone does not improve the efficiency of dose-finding trials either in terms of reducing the number of observed toxicities or improving the probability of correct dose selection. However, incorporating PK measures through a hierarchical model leads to better estimation of the dose-toxicity curve whilst maintaining the performance in terms of dose selection compared to dose-finding designs that do not incorporate PK information [25]. We developed an R package, dfpk, to provide a tool for physicians and statisticians involved in such clinical trials implementing the new method [26].

We developed and applied a novel dose-finding approach in the LEVNEONAT (NCT 02229123) trial that aims to find the optimal dose of Levetiracetam for reducing neonatal seizures with a maximum sample size of 50. In the trial, 3 primary outcomes were considered: efficacy and two types of toxicity that occur at the same time but can be measured earlier or later in time. The primary outcomes were modelled using a Bayesian approach with a logistic model for efficacy and a weighted likelihood with pseudo-outcomes for the two toxicities taking into account the correlation between the outcomes. This trial has received ethical committee approval and recruitment started in October 2017.

We have also focused on the development of possible extrapolation methods using information from studies in adults in the design of clinical trials in pediatrics. A unified approach for extrapolation and bridging adult information in early phase dose-finding studies was proposed. Using this approach we have investigated the choice of the dose range and calibration of prior density parameters of the dose-finding models for clinical trials involving children. The method uses adult observations, such as PK data, toxicity and efficacy. A large simulation study has shown that our method is robust and gives good performance in terms of dose selection [27, 28]. An R package, dfped, was developed to enable implementation of the new method [29].

In addition to developing methods to incorporate additional objective information in early phase trial design, we have also explored the possibility of incorporating subjective information such as expert opinion in a trial analysis. In particular, we have developed a method that reflects, when eliciting experts' opinions, how these depend on differences in experience, training and medical practice. The novel method proposed has been illustrated through a clinical trial comparing two treatments for idiopathic nephrotic syndrome, a rare disease in children (NCT 01092962). For each expert, a marginal prior was fitted from their elicitation of the distribution of treatment success. An overall prior was then constructed as a mixture of the individual physicians' priors using characteristics of the experts to weight their contribution in the mixture. A simulation study was used to evaluate several versions of the methodology [30].

## Meta-analysis and evidence synthesis methods in small population clinical trials

In order to survey the methodological challenges faced and the current practices applied in rare diseases, we performed systematic reviews of the literature in two exemplary rare indications, namely pediatric multiple sclerosis and Creutzfeldt-Jakob disease, focusing on design aspects, patient characteristics and statistical methodology. Our review yielded a total of 19 publications. While the quality of evidence appeared to be variable between the different fields, with mostly observational evidence in one and several randomized studies in the other, the design and analysis in most cases were based on standard techniques, suggesting that the use of more sophisticated statistical methods may contribute to some progress in these fields [31].

Meta-analysis methods are most commonly based on a normal model including variance components to account for estimation uncertainty as well as for potential heterogeneity between estimates [32]. We investigated this *normal-normal hierarchical model (NNHM)* with a focus on its performance and its limitations in the special case of only a few available estimates, and considering both classical and Bayesian approaches.

It is known that classical frequentist approaches to meta-analysis within the framework of the NNHM tend to run into problems when only few studies are available. We investigated the use of adjustments that had been proposed to ameliorate the poor behavior and found that a previously suggested modification of the common Hartung-Knapp-Sidik-Jonkman method performed better than other approaches especially in the common case of imbalanced study sizes [33].

A Bayesian approach offers another way to perform random-effect meta-analyses within the NNHM framework. One of the advantages is that the solution remains coherent also for small numbers of studies; on the other hand, careful prior specification is required, and the approach is usually computationally more demanding. We developed a general semi-analytical approach to solve the meta-analysis problem (and, in fact, a more general class of problems involving mixture distributions) via the `DIRECT` approach [34]. We have implemented this in the `bayesmeta` R package, to provide an efficient and user-friendly interface to Bayesian random-effects meta-analysis [35, 36]. The developed software allowed us to perform large-scale simulations to compare the different approaches in the special case of few studies; for an example of such a scenario, see Fig. 3. Here we could show that Bayesian methods perform well with respect to confidence/credible interval coverage and length [37, 38].

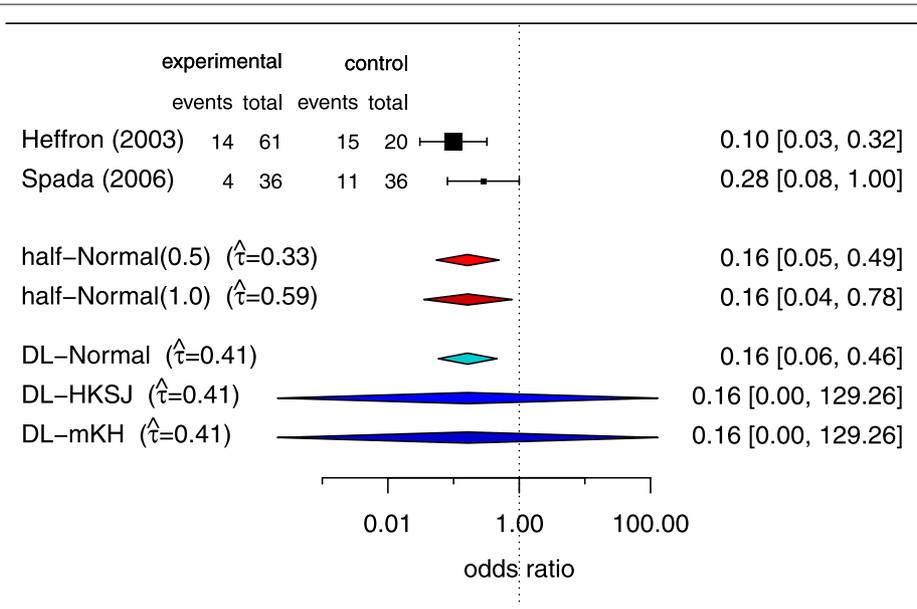

**Fig. 3** Meta-analyses of few studies are particularly challenging. Here, effect estimates from two studies in pediatric transplantation [51] are shown along with 5 different combined estimates based on several common approaches: two Bayesian analyses with different prior specifications, a normal approximation that is usually appropriate for large sample sizes, and two small-sample adjustments based on a Student-*t* distribution. We systematically investigated the long-run properties of popular meta-analysis procedures with a focus on few small studies [37, 38]. Figure reproduced from [38] under CC BY-NC-ND License [52]

The Bayesian model also allows implementation of a number of more advanced analysis strategies. We conducted further simulations to study different (arm-based and contrast-based) model variations in the special scenario of a single trial with available external evidence (Unkel, S., et al.: A Bayesian hierarchical framework for evidence synthesis for a single randomized controlled trial and observational data in small populations. In preparation.) motivated by an ongoing trial in Alport disease [39]. A series of studies may also be used to inform the analysis when the focus is not on an overall synthesis, but rather on a particular study that is to be viewed in the light of previously accumulated evidence. In this scenario, we investigated the use of shrinkage estimates to support data from a single trial in the light of external information [40].

Although a Bayesian approach holds promise for network meta-analysis, its considerable complexity hampers its general and easy application. We investigated the use of integrated nested Laplace approximations (INLA) to simplify and speed up computations, including continuous (normal) as well as count data (binomial) endpoints [41]. The implementation is available in the `nmaINLA` R package [42].

## Conclusions

Along with the Asterix and IDeAl projects, the InSPiRe project has provided substantial insights and further information to assist in clinical trial design for small patient populations, and to better inform regulators and decision-makers. Starting with a jointly-organised workshop, the three projects worked closely together both to pool expertise and to avoid overlapping research work. This paper has summarized the methodological work conducted as part of the InSPiRe project and referenced the main scientific publications where more details can be found. A summary of the project outputs in each of the methodological areas covered is given in Table 1. More details are available in the full project report [43].

In spite of the achievements of the InSPiRe, Asterix and IDeAl projects, the methodological work that can be completed in such relatively short-term projects is inevitably limited, with the move to widespread implementation of new methods in clinical trial practice extending well beyond the period of the projects themselves. This remains an area of ongoing work. The high level of regulation in clinical trials for evaluation of novel healthcare interventions, particularly novel medicinal products, means that application of our research results following the publication of innovative methodology often can occur only following dissemination to and engagement with regulatory authorities. A major regulatory development of relevance to clinical trials in small populations during the time of the InSPiRe project has been the production of the draft EMA PDCO Reflection paper on extrapolation of efficacy and safety in pediatric medicine development [44]. Following the publication of the EMA Concept paper, a workshop of an EMA Extrapolation expert group was held in September 2015, leading to the production of the draft Reflection paper in March 2016 [45] and a public workshop held by EMA in May 2016. InSPiRe team members have been fully involved in these meetings and in development of these drafts along with colleagues from the Asterix and IDeAl projects. The level of interest in and commitment to the InSPiRe, Asterix and IDeAl projects by the EMA is also demonstrated by their hosting of a joint meeting of the three projects in March 2017. Along with the coordinators of the Asterix and IDeAl projects, members of the InSPiRe team also joined the Steering Committee of the Small-populations Clinical Trial Task Force of the International Rare Diseases Research Consortium (IRDiRC). The task force produced a report of their recommendations at a workshop held at EMA in March 2016 [46]. Together with colleagues from the Asterix and IDeaAl projects, members of the InSPiRe team have also contributed to ongoing regulatory discussions on data sharing [47].

Besides issues of regulatory harmonization, another hurdle to the widespread implementation of novel statistical methods is the availability of software. To address this issue, we have produced open access statistical software to run on the freely available software environment R [48] to implement the new approaches that we have developed in meta-analysis and network meta-analysis (packages `bayesmeta` and `nmaINLA`) and in dose-finding (package `dfpk` and `dfped`). These software packages are available for download from the Comprehensive R Archive Network (https://cran.r-project.org).

**Abbreviations**
Asterix: Advances in small trials design for regulatory innovation and excellence; EMA: European medicines agency; EU FP7: European union's seventh framework programme for research, technological development and demonstration; IDeAl: Integrated design and analysis of small population group trials; INLA: Integrated nested Laplace approximations; InSPiRe: Innovative methodology for small populations research; IRDiRC: International rare diseases research consortium; NNHM: Normal-normal hierarchical model; PD: Pharmacodynamic(s); PDCO: Pediatric committee; PK: Pharmacokinetic(s); RCT: Randomized controlled trial; VOI: Value-of-information

**Acknowledgements**
We are grateful to two reviewers for their helpful comments.

**Funding**
This work was conducted as part of the InSPiRe (Innovative methodology for small populations research) project funded by the European Union's Seventh Framework Programme for research, technological development and demonstration under grant agreement number FP HEALTH 2013–602144.

## Authors' contributions

NS, TF, MP and SZ contributed to the conduct, conception and design of the research and the drafting of the manuscript. All authors contributed to the conduct of the research and have read and approved the final manuscript.

## Ethics approval and consent to participate

Not applicable.

## Consent for publication

Not applicable.

## Competing interests

The authors declare that they have no competing interests.

# Publisher's Note

Springer Nature remains neutral with regard to jurisdictional claims in published maps and institutional affiliations.

## Author details

[1]UMG, Göttingen, Germany. [2]Section of Medical Statistics, CeMSIIS, Medical University of Vienna, Vienna Austria. [3]INSERM, U1138, team 22, Centre de Recherche des Cordeliers, Université Paris 5, Université Paris 6, Paris, France. [4]INSERM, Hôpital Robert-Debré, APHP, University Paris 7, Paris, France. [5]BfArM, Bonn, Germany. [6]INSERM, IAME, UMR 1137, Univ Paris Diderot, Sorbonne Paris Cité, Paris, France. [7]INSERM, CIC 1414, Univ Rennes-1, Rennes, France. [8]Clinical Trials Consulting and Training Limited, Buckingham, UK. [9]Mediana Inc., Overland Park, KS, USA. [10]Warwick Medical School, University of Warwick, Coventry, UK. [11]Department of Statistics, Stockholm University, Stockholm Sweden. [12]Complexity Science, University of Warwick, Coventry, UK.